\def\be{\begin{equation}}
\def\ee{\end{equation}}
\def\ba{\begin{array}}

\def\ea{\end{array}}

\documentclass[12pt]{article}
\usepackage{amsfonts,amssymb,amsmath,graphicx}
\newtheorem{theorem}{Theorem}

\begin{document}
\parskip=3pt
\parindent=18pt
\baselineskip=20pt
\setcounter{page}{1}
\centerline{\large\bf Notes on modified trace distance measure of coherence}
\vspace{6ex}
\centerline{{\sf Bin Chen,$^{\star}$}
\footnote{\sf Corresponding author: chenbin5134@163.com}
~~~ {\sf Shao-Ming Fei$^{\natural,\sharp}$}
}
\vspace{4ex}
\centerline
{\it $^\star$ School of Mathematical Sciences, Tianjin Normal University, Tianjin 300387, China}\par
\centerline
{\it $^\natural$ School of Mathematical Sciences, Capital Normal University, Beijing 100048, China}\par
\centerline
{\it $^\sharp$ Max-Planck-Institute for Mathematics in the Sciences, 04103 Leipzig, Germany}\par
\vspace{6.5ex}
\parindent=18pt
\parskip=5pt
\begin{center}
\begin{minipage}{5in}
\vspace{3ex}
\centerline{\large Abstract}
\vspace{4ex}
We investigate the modified trace distance measure of coherence recently introduced in [Phys. Rev. A 94, 060302(R) (2016)].
We show that for any single-qubit state, the modified trace norm of coherence is equal to the $l_{1}$-norm of coherence.
For any $d$-dimensional quantum system, an analytical formula of this measure for a class of maximally coherent mixed states is provided.
The trade-off relation between the coherence quantified by the new measure and the mixedness quantified by the trace norm is also discussed.
Furthermore, we explore the relation between the modified trace distance measure of coherence and other measures
such as the $l_{1}$-norm of coherence and the geometric measure of coherence.
\end{minipage}
\end{center}

\newpage

\section{Introduction}

Quantum coherence, as one of the most fundamental and striking features in quantum mechanics, plays a key role in various research fields such as low-temperature thermodynamics \cite{y1,y2,y3,y4,y5}, quantum biology \cite{y6,y7,y8,y9,y10,y11}, nanoscale physics \cite{y12,y13}, ect.
Although characterizing coherence in a rigorous framework is a desirable and intriguing task, considerable progresses have only been made in quantum optics \cite{y14,y15}.
Recently, Baumgratz \emph{et al.} proposed a resource-theoretic framework for quantifying quantum coherence \cite{Baumgratz} by following the quantitative theory of quantum entanglement.
This framework includes two important concepts -- incoherent states and incoherent operations, which are analogous to the separable states and local operations and classical communication (LOCC) respectively in the quantum entanglement theory.
A nonnegative function $C$ defined on a space of quantum states can be used as a measure of coherence, if it satisfies the following four conditions \cite{Baumgratz}:

(B1) $C(\rho)=0$ if and only if $\rho$ is an incoherent state, i.e., $\rho$ can be written as
$\rho=\sum_{i=1}^{d}p_{i}|i\rangle\langle i|,~p_{i}\geq0,~\sum_{i=1}^{d}p_{i}=1$,
for a fixed basis $\{|i\rangle\}_{i=1}^{d}$ in a $d$-dimensional Hilbert space;

(B2) $C(\Lambda(\rho))\leq C(\rho)$ for any incoherent operation $\Lambda$, i.e., $\Lambda$ is a completely positive trace preserving (CPTP) map,
$\Lambda(\rho)=\sum_{n}K_{n}\rho K_{n}^{\dag}$, where $\{K_{n}\}$ is a set of incoherent Kraus operators satisfying
$K_{n}\mathcal{I}K_{n}^{\dag}\subseteq\mathcal{I}$ for all $n$,
$\mathcal{I}$ is the set of incoherent states;

(B3) $\sum_{n}p_{n}C(\rho_{n})\leq C(\rho)$, where $p_{n}=\mathrm{Tr}(K_{n}\rho K_{n}^{\dag}),\rho_{n}=K_{n}\rho K_{n}^{\dag}/p_{n},\{K_{n}\}$ is a set of incoherent Kraus operators;

(B4) $C$ is a  convex function, i.e., $C(\sum_{i}p_{i}\rho_{i})\leq\sum_{i}p_{i}C(\rho_{i})$ for any set of quantum states $\{\rho_{i}\}$ and any probability distribution $\{p_{i}\}$.

Conditions (B2) and (B3) are often referred as monotonicity and strong monotonicity under incoherent channel, respectively.
Similar to the entanglement quantifiers, very few functions satisfy the strong monotonicity condition and can be used as proper measures of coherence \cite{rmp}.
In their seminal paper, Baumgratz \emph{et al.} presented two coherence measures satisfying conditions (B1)--(B4) for all states,
i.e., the $l_{1}$-norm-based measure $C_{l_{1}}$ defined as $C_{l_{1}}(\rho)=\min_{\sigma\in\mathcal{I}}\|\rho-\sigma\|_{l_{1}}=\sum_{i\neq j}|\rho_{ij}|$,
and the relative entropy based measure $C_{r}$ defined as $C_{r}(\rho)=\min_{\sigma\in\mathcal{I}}S(\rho\|\sigma)$.
Then a natural question is raised that whether $l_{p}$-norm and Schatten-$p$-norm can also be employed to be proper coherence measures for all $p$.
However, this is not the case.
In Ref. \cite{Rana} it has been shown that both $l_{p}$-norm-based and Schatten-$p$-norm-based measures violate the strong monotonicity for $p>1$,
leaving the problem whether the trace norm (Schatten-$1$-norm) of coherence is a legitimate coherence measure still open \cite{wz}.

Recently, Yu \emph{et al.} put forward an alternative framework for quantifying coherence \cite{Tdm}.
This framework consists of the following three conditions that a function $C$ should satisfy as a measure of coherence:

(C1) $C(\rho)\geq0$, and $C(\rho)=0$ if and only if $\rho\in\mathcal{I}$;

(C2) $C(\Lambda(\rho))\leq C(\rho)$ for any incoherent operation $\Lambda$;

(C3) $C(p_{1}\rho_{1}\oplus p_{2}\rho_{2})=p_{1}C(\rho_{1})+p_{2}C(\rho_{2})$ for block diagonal states $\rho$ in the incoherent basis.

It has been shown that this framework is equivalent to the previous one proposed by Baumgratz \emph{et al.}.
That is to say, the three conditions (C1)--(C3) can be derived from conditions (B1)--(B4), and vice visa.
However, the new framework is more convenient for various applications.
By using condition (C3), the authors in Ref. \cite{Tdm} proved that the trace norm of coherence is not a legitimate coherence measure, thus it must violate the strong monotonicity.
Furthermore, they introduced a new measure, called the modified trace norm of coherence,
\begin{equation}\label{mt}
C_{\mathrm{tr}}^{\prime}(\rho)=\min_{\lambda\geq0,\delta\in\mathcal{I}}\|\rho-\lambda\delta\|_{\mathrm{tr}},
\end{equation}
which can be shown to satisfy (C1)--(C3).
Like the $l_{1}$-norm-based measure $C_{l_{1}}$ and the relative entropy based measure $C_{r}$, this new measure $C_{\mathrm{tr}}^{\prime}$ is also worthy of being further investigated.

In this paper, we compute the modified trace norm of coherence for any single-qubit state and a class of maximally coherent mixed states for any qudit system.
We also discuss the trade-off relations between the coherence quantified by the modified trace distance measure and the mixedness quantified by the trace norm.
Moreover, we study the relations among the modified trace distance measure, the $l_{1}$-norm-based measure and the geometric measure of coherence.

\section{Modified trace norm of coherence}
In this paper, we fix a set of basis $\{|i\rangle\}_{i=0}^{d-1}$ in a $d$-dimensional Hilbert space.
To find the analytic form of modified trace norm of coherence defined in (\ref{mt}) for any single-qubit state,
let us consider a 2$\times$2 Hermitian matrix $A$ with eigenvalues $\lambda_{1}$ and $\lambda_{2}$.
We have that $\|A\|_{\mathrm{tr}}^{2}=(|\lambda_{1}|+|\lambda_{2}|)^{2}=\mathrm{Tr}(A^{2})+2|\mathrm{det}(A)|$.
Using this fact, we have the following result.

\emph{Proposition 1.} Let $\rho=\frac{1}{2}(I+\overrightarrow{r}\cdot\overrightarrow{\sigma})$ be a qubit state, where
$\overrightarrow{r}=(r_{1},r_{2},r_{3}),~\overrightarrow{\sigma}=(\sigma_{1},\sigma_{2},\sigma_{3})$.
Then the modified trace norm of coherence for $\rho$ is equal to the $l_{1}$-norm of coherence, i.e.,
$C_{\mathrm{tr}}^{\prime}(\rho)=C_{l_{1}}(\rho)=\sqrt{r_{1}^{2}+r_{2}^{2}}$.

\emph{Proof.} We need to minimize $\|\rho-\lambda\delta\|_{\mathrm{tr}}$, where $\lambda\geq0,~\delta=\mathrm{diag}\{x_{1},x_{2}\},~x_{1},x_{2}\geq0,~x_{1}+x_{2}=1$.
Under the fixed basis, one has
\begin{equation*}
\rho-\lambda\delta=\begin{pmatrix}
\frac{1}{2}(1+r_{3})-\lambda x_{1} & \frac{1}{2}(r_{1}-\mathrm{i}r_{2})\\\frac{1}{2}(r_{1}+\mathrm{i}r_{2}) & \frac{1}{2}(1-r_{3})-\lambda x_{2}
\end{pmatrix}.
\end{equation*}
Therefore
\begin{equation}
\begin{split}
\|\rho-\lambda\delta\|_{\mathrm{tr}}^{2}=&\left[\frac{1}{2}(1+r_{3})-\lambda x_{1}\right]^{2}+\left[\frac{1}{2}(1-r_{3})-\lambda x_{2}\right]^{2}+\frac{1}{2}(r_{1}^{2}+r_{2}^{2})\\
&+2\left|\left[\frac{1}{2}(1+r_{3})-\lambda x_{1}\right]\left[\frac{1}{2}(1-r_{3})-\lambda x_{2}\right]-\frac{1}{4}(r_{1}^{2}+r_{2}^{2})\right|.
\end{split}
\end{equation}
Set $a=\frac{1}{2}(1+r_{3})-\lambda x_{1},~b=\frac{1}{2}(1-r_{3})-\lambda x_{2},~c=\frac{1}{4}(r_{1}^{2}+r_{2}^{2})$.
Using the inequality \cite{Rana}: $a^{2}+b^{2}+2|ab-c|\geq2|c|$, we get
\begin{equation}
\|\rho-\lambda\delta\|_{\mathrm{tr}}\geq\sqrt{r_{1}^{2}+r_{2}^{2}},
\end{equation}
and the equality holds when $\lambda=1,~\delta=\mathrm{diag}\{\frac{1}{2}(1+r_{3}),\frac{1}{2}(1-r_{3})\}$.
Thus we have $C_{\mathrm{tr}}^{\prime}(\rho)=\sqrt{r_{1}^{2}+r_{2}^{2}}$.
On the other hand, $C_{l_{1}}(\rho)=\frac{1}{2}|r_{1}-\mathrm{i}r_{2}|+\frac{1}{2}|r_{1}+\mathrm{i}r_{2}|=\sqrt{r_{1}^{2}+r_{2}^{2}}$.
Hence $C_{\mathrm{tr}}^{\prime}(\rho)=C_{l_{1}}(\rho)$ holds for any qubit state. \quad $\Box$

For high-dimensional quantum states, the computation of $C_{\mathrm{tr}}^{\prime}(\rho)$ becomes more difficult, since it is not easy to find the closest incoherent state with multiplier $\lambda$.
However, we can calculate the modified trace norm of coherence for a class of important coherent states -- the maximally coherent mixed states
(MCMS) \cite{Singh}, which are defined as
\begin{equation}\label{MCMS}
\rho_{m}=p|\phi_{d}\rangle\langle\phi_{d}|+\frac{1-p}{d}\mathbb{I}_{d},
\end{equation}
where $0<p\leq1$, and $|\phi_{d}\rangle=\frac{1}{\sqrt{d}}\sum_{i=0}^{d-1}|i\rangle$ is the maximally coherent state.
To this end, let us consider a class of unitary operators $U_{n}=\sum_{k=0}^{d-1}|k\oplus n\rangle\langle k|,~n=0,\ldots d-1$,
where $k\oplus n$ denotes $(k+n)~\mathrm{mod}~d$.
Taking into account the unitary invariance and the subadditivity of the trace norm, we have
\begin{equation}
\begin{split}
\|\rho_{m}-\lambda\delta\|_{\mathrm{tr}}&=\frac{1}{d}\sum_{n=0}^{d-1}\left\|U_{n}\left(p|\phi_{d}\rangle\langle\phi_{d}|+\frac{1-p}{d}\mathbb{I}_{d}-
\lambda\delta\right)U_{n}^{\dagger}\right\|_{\mathrm{tr}}\\
&\geq\frac{1}{d}\left\|\sum_{n=0}^{d-1}U_{n}\left(p|\phi_{d}\rangle\langle\phi_{d}|+\frac{1-p}{d}\mathbb{I}_{d}-
\lambda\delta\right)U_{n}^{\dagger}\right\|_{\mathrm{tr}}.
\end{split}
\end{equation}
By using the facts that $U_{n}|\phi_{d}\rangle=|\phi_{d}\rangle,~0\leq n\leq d-1$, and $\sum_{n=0}^{d-1}U_{n}\delta U_{n}^{\dagger}=\mathbb{I}_{d}$ \cite{Tdm},
we get
\begin{equation}
\begin{split}
\|\rho_{m}-\lambda\delta\|_{\mathrm{tr}}&\geq\left\|p|\phi_{d}\rangle\langle\phi_{d}|+\frac{1-p-\lambda}{d}\mathbb{I}_{d}\right\|_{\mathrm{tr}}\\
&=\left|p+\frac{1-p-\lambda}{d}\right|+(d-1)\left|\frac{1-p-\lambda}{d}\right|\\
&=|p+x|+(d-1)|x|,
\end{split}
\end{equation}
where $x=\frac{1-p-\lambda}{d}$, and $x\leq\frac{1-p}{d}$ since $\lambda\geq0$.
To find the minimum value of $|p+x|+(d-1)|x|$, we consider the following cases:

(i) If $x\leq-p$, then $|p+x|+(d-1)|x|=-p-dx\geq(d-1)p\geq p$;

(ii) If $-p\leq x\leq0$, then $|p+x|+(d-1)|x|=p-(d-2)x\geq p$;

(iii) If $0\leq x\leq\frac{1-p}{d}$, then $|p+x|+(d-1)|x|=p+dx\geq p$.\\
Thus we have $\|\rho_{m}-\lambda\delta\|_{\mathrm{tr}}\geq p$.
Setting $\lambda=1-p$ and $\delta=\frac{1}{d}\mathbb{I}_{d}$, one can easily get $\|\rho_{m}-\lambda\delta\|_{\mathrm{tr}}=p$.
Therefore we obtain $C_{\mathrm{tr}}^{\prime}(\rho_{m})=p$.

\section{Trade-off relations between coherence and mixedness}
In Ref. \cite{Singh}, Singh \emph{et al.} discussed the relations between quantum coherence and mixedness in any $d$-dimensional quantum system.
They claimed that the amount of coherence that a quantum system can possess must be limited, if the mixedness of the system is fixed.
A trade-off relation between coherence quantified by the $l_{1}$-norm $C_{l_{1}}(\rho)$ and the mixedness quantified by the normalized linear entropy
$M_{l}(\rho)=\frac{d}{d-1}(1-\mathrm{Tr}(\rho^{2}))$ is given \cite{Singh}:
\begin{equation}\label{cm0}
\frac{C_{l_{1}}^{2}(\rho)}{(d-1)^{2}}+M_{l}(\rho)\leq1.
\end{equation}
Quantum states with maximal coherence for a fixed mixedness are called maximally coherent mixed states (MCMS).
It has been shown that $\rho_{m}$ defined in (\ref{MCMS}) is the only form of MCMS with respect to the above inequality \cite{Singh}.
However, for other measures of coherence and mixedness, few results have been known so far for such trade-off relations \cite{Singh,Zhang}.

In this section, we discuss trade-off relations between coherence quantified by the modified trace distance measure and the mixedness quantified by the trace norm.
To define a new measure of quantum mixedness, we first determine the range of $\|\rho-\frac{1}{d}\mathbb{I}_{d}\|_{\mathrm{tr}}$ for arbitrary qudit state $\rho$.
Let $\lambda_{i},~i=1,\ldots, d$, be the eigenvalues of $\rho$, and assume that
$\lambda_{1}\geq\lambda_{2}\geq\cdots\geq\lambda_{k}\geq\frac{1}{d}\geq\lambda_{k+1}\geq\cdots\geq\lambda_{d},~1\leq k\leq d$.
Then we have
\begin{equation}
\begin{split}
\left\|\rho-\frac{1}{d}\mathbb{I}_{d}\right\|_{\mathrm{tr}}&=\sum_{i=1}^{d}\left|\lambda_{i}-\frac{1}{d}\right|\\
&=\sum_{i=1}^{k}\lambda_{i}-\frac{k}{d}-\left(\sum_{i=k+1}^{d}\lambda_{i}-\frac{d-k}{d}\right)\\
&=2\left(\sum_{i=1}^{k}\lambda_{i}-\frac{k}{d}\right)\\
&\leq2\left(1-\frac{1}{d}\right),
\end{split}
\end{equation}
where the equality holds if $\rho$ is a pure state.
Thus we can define a new measure of mixedness based on the trace norm
\begin{equation}
M_{\mathrm{tr}}(\rho)=1-\frac{d}{2(d-1)}\left\|\rho-\frac{1}{d}\mathbb{I}_{d}\right\|_{\mathrm{tr}}.
\end{equation}
It can easily be seen that $0\leq M_{\mathrm{tr}}(\rho)\leq1$, and $M_{\mathrm{tr}}(\rho)=0$ if $\rho$ is pure.

For general cases, the trade-off relations between $C_{\mathrm{tr}}^{\prime}(\rho)$ and $M_{\mathrm{tr}}(\rho)$ are very difficult to derive,
since we have no analytical form of $C_{\mathrm{tr}}^{\prime}(\rho)$ for arbitrary quantum states.
However, in the next, we will present such trade-off relation for single-qubit state and a complementarity relation for $\rho_{m}$.

Consider again the single-qubit state $\rho=\frac{1}{2}(I+\overrightarrow{r}\cdot\overrightarrow{\sigma})$ with eigenvalues
$\lambda_{1}=\frac{1}{2}(1+|\overrightarrow{r}|)$ and $\lambda_{2}=\frac{1}{2}(1-|\overrightarrow{r}|)$.
Then we have $M_{\mathrm{tr}}(\rho)=1-(|\lambda_{1}-\frac{1}{2}|+|\lambda_{2}-\frac{1}{2}|)=1-|\overrightarrow{r}|$.
Using the result derived in Proposition 1, $C_{\mathrm{tr}}^{\prime}(\rho)=\sqrt{r_{1}^{2}+r_{2}^{2}}$, we have the following trade-off relation,
\begin{equation}
C_{\mathrm{tr}}^{\prime}(\rho)+M_{\mathrm{tr}}(\rho)=\sqrt{r_{1}^{2}+r_{2}^{2}}+(1-|\overrightarrow{r}|)\leq1.
\end{equation}
The equality holds if and only if $r_{3}=0$.
Thus the maximally coherent mixed states in this case are those states with Bloch vectors in the closed unit disk on the $r_{1}$-$r_{2}$ plane.

For a class of MCMS $\rho_{m}$ given in (\ref{MCMS}), one can easily get $\|\rho_{m}-\frac{1}{d}\mathbb{I}_{d}\|_{\mathrm{tr}}=\frac{2p(d-1)}{d}$ by simple algebra, thus $M_{\mathrm{tr}}(\rho_{m})=1-p$.
Taking into account that $C_{\mathrm{tr}}^{\prime}(\rho_{m})=p$, we obtain the following complementarity relation between coherence and mixedness
\begin{equation}
C_{\mathrm{tr}}^{\prime}(\rho_{m})+M_{\mathrm{tr}}(\rho_{m})=1.
\end{equation}

\section{Relation among $C_{\mathrm{tr}}^{\prime},~C_{l_{1}}$ and the geometric measure of coherence}
The $l_{1}$ norm of coherence has been deeply investigated and the analytical expression of $C_{l_{1}}$ is given in \cite{Baumgratz}.
It is interesting to find the relation among $C_{l_{1}}$ and other measures.
The interrelations between $C_{l_{1}}$ and the relative entropy of coherence $C_{r}$ have been derived in \cite{Rana}.
Here we will prove that $C_{l_{1}}$ is an upper bound for the modified trace norm of coherence $C_{\mathrm{tr}}^{\prime}$.

In Ref. \cite{Streltsov}, the authors introduced the geometric measure of coherence
defined by $C_{g}(\rho)=1-\max_{\sigma\in\mathcal{I}}F(\rho,\sigma)$,
where $F(\rho,\sigma)=(\mathrm{Tr}\sqrt{\sqrt{\sigma}\rho\sqrt{\sigma}})^{2}$ is the fidelity of two density operators $\rho$ and $\sigma$.
The computation of $C_{g}$ for any qudit state is formidably difficult.
Recently Zhang \emph{et al.} provided the lower and upper bounds of $C_{g}$ \cite{Zhang}.
It is also worth studying the relations between $C_{g}$ and $C_{\mathrm{tr}}^{\prime}$.

We have the following Theorem.
\begin{theorem}
Let $\rho$ be a qudit state. Then we have
\begin{equation}
C_{g}(\rho)\leq C_{\mathrm{tr}}^{\prime}(\rho)\leq C_{l_{1}}(\rho).
\end{equation}
\end{theorem}

\emph{Proof.} To see that $C_{l_{1}}(\rho)$ is an upper bound of $C_{\mathrm{tr}}^{\prime}(\rho)$, we only need to use the fact that $\|A\|_{\mathrm{tr}}\leq\|A\|_{l_{1}}$ for any Hermitian matrix $A$ \cite{ZX}.
Then we have
\begin{equation}
\begin{split}
C_{\mathrm{tr}}^{\prime}(\rho)&=\min_{\lambda\geq0,\delta\in\mathcal{I}}\|\rho-\lambda\delta\|_{\mathrm{tr}}\\
&\leq\min_{\delta\in\mathcal{I}}\|\rho-\delta\|_{\mathrm{tr}}\\
&\leq\min_{\delta\in\mathcal{I}}\|\rho-\delta\|_{l_{1}}\\
&=C_{l_{1}}(\rho).
\end{split}
\end{equation}

Now we prove that $C_{\mathrm{tr}}^{\prime}(\rho)\geq C_{g}(\rho)$.
For two positive semidefinite matrices $P$ and $Q$, it holds that $\|P-Q\|_{\mathrm{tr}}\geq\|\sqrt{P}-\sqrt{Q}\|_{\mathrm{HS}}^{2}$,
where $\|A\|_{\mathrm{HS}}=\sqrt{\mathrm{Tr}(A^{\dag}A)}$ is the Hilbert-Schmidt norm \cite{Watrous}.
Then we have $C_{\mathrm{tr}}^{\prime}(\rho)\geq\min_{\lambda\geq0,\delta\in\mathcal{I}}\|\sqrt{\rho}-\sqrt{\lambda\delta}\|_{\mathrm{HS}}^{2}$.
Thus we only need to minimize $\|\sqrt{\rho}-\sqrt{\lambda\delta}\|_{\mathrm{HS}}^{2}$.
Let $\sqrt{\rho}=\sum_{i,j}b_{ij}|i\rangle\langle j|,~\delta=\sum_{i}x_{i}|i\rangle\langle i|,~x_{i}\geq0,~\sum_{i}x_{i}=1$.
Note that
\begin{equation}
\begin{split}
\|\sqrt{\rho}-\sqrt{\lambda\delta}\|_{\mathrm{HS}}^{2}&=\mathrm{Tr}(\sqrt{\rho}-\sqrt{\lambda\delta})^{2}\\
&=\mathrm{Tr}(\rho+\lambda\delta)-2\mathrm{Tr}(\sqrt{\rho}\sqrt{\lambda\delta})\\
&=1+\lambda-2\sum_{i}b_{ii}\sqrt{\lambda x_{i}}\\
&=1-\sum_{i}b_{ii}^{2}+\sum_{i}(b_{ii}-\sqrt{\lambda x_{i}})^{2}\\
&\geq1-\sum_{i}b_{ii}^{2},
\end{split}
\end{equation}
and the equality holds if and only if $\lambda x_{i}=b_{ii},~\forall i$, which yields that $\lambda=\sum_{i}b_{ii}^{2},~x_{i}=b_{ii}^{2}/\sum_{i}b_{ii}^{2}$.
Then we obtain $C_{\mathrm{tr}}^{\prime}(\rho)\geq1-\sum_{i}b_{ii}^{2}$.
On the other hand, it has been shown that $C_{g}(\rho)\leq1-\sum_{i}b_{ii}^{2}$ \cite{Zhang}.
Therefore we have $C_{\mathrm{tr}}^{\prime}(\rho)\geq C_{g}(\rho)$.
This completes the proof.  \quad $\Box$

\begin{figure}
\centering
\includegraphics[width=7cm]{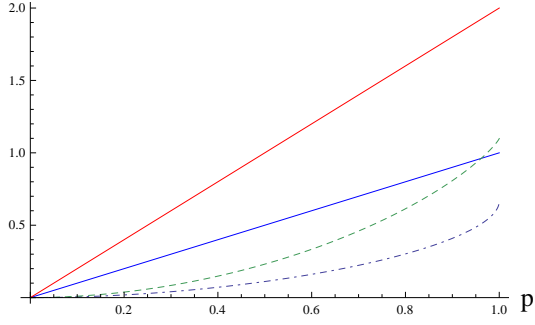}
\caption{The red and blue solid lines are the values of $C_{l_{1}}(\rho_{m})$ and $C_{\mathrm{tr}}^{\prime}(\rho_{m})$, respectively.
The dot-dashed line is $C_{g}(\rho_{m})$, and the dashed line is $C_{r}(\rho_{m})$.}\label{tr}
\end{figure}

As an example, FIG \ref{tr} shows the relations among $C_{\mathrm{tr}}^{\prime}(\rho_{m})$, $C_{g}(\rho_{m})$, $C_{l_{1}}(\rho_{m})$ and $C_{r}(\rho_{m})$ for $d=3$.
It can be seen that $C_{\mathrm{tr}}^{\prime}(\rho_{m})>C_{r}(\rho_{m})$ when $0<p<0.96151$, and $C_{\mathrm{tr}}^{\prime}(\rho_{m})\leq C_{r}(\rho_{m})$ when $0.96151\leq p\leq1$.
Thus there is no order relation between the quantities $C_{\mathrm{tr}}^{\prime}(\rho)$ and $C_{r}(\rho)$ in general.

\section{Conclusion}
We have investigated the modified trace norm of coherence $C_{\mathrm{tr}}^{\prime}(\rho)$.
The analytical formulae of this new measure are provided for any single-qubit state and for a class of maximally coherent mixed states.
For qubit states, the modified trace norm of coherence is equal to the $l_{1}$-norm of coherence.
However, the calculation of $C_{\mathrm{tr}}^{\prime}(\rho)$ for arbitrary quantum states is more complicated,
since it is not easy to find the optimal incoherent state $\delta$ and the multiplier $\lambda$ simultaneously in general.

We have also discussed the trade-off relation between coherence quantified by this new measure and the mixedness quantified by the trace norm.
A new class of maximally coherent mixed states for qubit system has been obtained by providing a trade-off relation between these two quantities.
For a special class of coherent states given in (\ref{MCMS}), a complementarity relation has also been presented.

As a new measure of coherence, its relations to other measures are worth studying.
We have shown that the $l_{1}$-norm coherence provides an upper bound for $C_{\mathrm{tr}}^{\prime}(\rho)$,
while the geometric measure of coherence $C_{g}(\rho)$ is the lower bound of $C_{\mathrm{tr}}^{\prime}(\rho)$.
Further efforts should be made toward analytical formulae of $C_{\mathrm{tr}}^{\prime}(\rho)$ for arbitrary or any other special classes of states $\rho$,
new trade-off relations between the coherence and the mixedness,
as well as the relations to other coherence quantifiers.

\vspace{2.5ex}
\noindent{\bf Acknowledgments}\, \,
This work is supported by the NSF of China under Grant No. 11675113.

\end{document}